# On the Achievable Rate of the Fading Dirty Paper Channel with Imperfect CSIT

Chinmay S. Vaze and Mahesh K. Varanasi

*Abstract*—The problem of dirty paper coding (DPC) over the (multi-antenna) fading dirty paper channel (FDPC) $Y = H(X + S) + Z$ is considered when there is imperfect knowledge of the channel state information $H$ at the transmitter (CSIT). The case of FDPC with positive definite (p.d.) input covariance matrix was studied by the authors in a recent paper, and here the more general case of positive semi-definite (p.s.d.) input covariance is dealt with. Towards this end, the choice of auxiliary random variable is modified. The algorithms for determination of inflation factor proposed in the p.d. case are then generalized to the case of p.s.d. input covariance. Subsequently, the largest DPC-achievable high-SNR (signal-to-noise ratio) scaling factor over the no-CSIT FDPC with p.s.d. input covariance matrix is derived. This scaling factor is seen to be a non-trivial generalization of the one achieved for the p.d. case. Next, in the limit of low SNR, it is proved that the choice of all-zero inflation factor (thus treating interference as noise) is optimal in the 'ratio' sense, regardless of the covariance matrix used. Further, in the p.d. covariance case, the inflation factor optimal at high SNR is obtained when the number of transmit antennas is greater than the number of receive antennas, with the other case having been already considered in the earlier paper. Finally, the problem of joint optimization of the input covariance matrix and the inflation factor is dealt with, and an iterative numerical algorithm is developed.

## I. INTRODUCTION

IN this paper, we continue with our study of the problem of dirty paper coding (DPC) [1] over the fading dirty paper channel (FDPC) $Y = H(X + S) + Z$ that was initiated recently in [2]. We consider the case in which there is imperfect and perfect knowledge of the fading process $H$ at the transmitter (CSIT) and the receiver (CSIR), respectively. This problem is important from the point of view of application of DPC to the Gaussian multiple-input multiple-output (MIMO) broadcast channel (BC) with imperfect CSIT. In [2], we considered the case of input covariance matrix being positive definite (p.d.). Those results are restrictive in a sense that if a beamforming and DPC-based transmission scheme is to be used over the Gaussian MIMO BC (beamforming-based schemes are popularly being studied for the BC) then the results of [2] alone will not suffice. With this motivation, this paper expands the scope to include a more general case of positive semi-definite (p.s.d.) input covariance.

The most crucial part of the problem of DPC over the imperfect-CSIT FDPC problem is to determine the optimal inflation factor. Towards this end, for the case of p.d. input covariance, we proposed two iterative, numeri-

This work was supported in part by NSF Grants CCF-0431170 and CCF-0728955. The authors are with the Department of Electrical and Computer Engineering, University of Colorado, Boulder, CO 80309-0425 USA (e-mail: Chinmay.Vaze, varanasi@colorado.edu).

cal algorithms. These algorithms are seen to perform well. However, these algorithms can not be used directly when the input covariance matrix is p.s.d. This is because the achievable rate expression derived there is valid only when the covariance matrix is p.d. We thus slightly modify the choice of auxiliary random variable (RV) (see [3], [1] for definition), derive the achievable-rate expression for the p.s.d. case, and then generalize the previously-developed algorithms to apply to the case of current interest. Next, in Section IV, we derive the optimal high-SNR scaling factor achievable over the FDPC with p.s.d. input covariance. This scaling factor has several interesting characteristics that are not observed in the p.d. case. Further, an important result regarding the low-SNR behavior of the achievable rate is proved. Numerical results obtained using our algorithms are furnished in Section V.

Finally, we consider the problem of optimizing the achievable rate of the no-CSIT FDPC over the input covariance matrix. Such a problem is well-studied for the MIMO channel [4], [5], [6]. The problem becomes more interesting in the case of FDPC because of the dependence of the inflation factor on the input covariance matrix. As a result, the covariance optimization must now be treated as a problem of joint optimization over the inflation factor and the input covariance. We develop an iterative algorithm for this joint optimization in Section VI.

<u>Notation:</u> For a square matrix $A$, $\text{tr}(A)$, $\text{rank}(A)$, and $|A|$ denote its trace, rank, and determinant, respectively. For any general matrix $A$, $A^*$, $A^+$, $\text{null}(A)$ denote its complex-conjugate transpose, pseudo-inverse, and null space, respectively. $I_m$ is $m \times m$ identity matrix. $\mathbb{E}_X(\cdot)$ denotes expectation over RV $X$.

## II. CHANNEL MODEL

The $t \times r$ FDPC is defined via equation $Y = H(X + S) + Z$. Here, $X \sim \mathcal{CN}(0, \Sigma_X)$ is the transmitted signal with the power constraint of $\text{tr}(\Sigma_X) \leq P$; the interference $S \sim \mathcal{CN}(0, \Sigma_S)$ is assumed to be known non-causally at the transmitter but not at the receiver; $Z \sim \mathcal{CN}(0, \Sigma_Z)$ is the additive noise; and $X$, $S$, and $Z$ are independent. The channel matrix $H$ is assumed to be known perfectly at the receiver, whereas at the transmitter, only an estimate $\hat{H}$ is known. Assume $|\Sigma_Z| > 0$. Let $\text{tr}(\Sigma_S) = Q$, $\text{tr}(\Sigma_Z) = N$; and define SNR $= \frac{P}{N}$.

Let the rank of $\Sigma_X$ be $m \leq t$. Then we can write $\Sigma_X = TT^*$ for some $t \times m$ matrix $T$, and therefore, let $X = TX'$ for some $X' \sim \mathcal{CN}(0, I_m)$. We thus choose the auxiliary RV as $U = X' + WS$, as opposed to the choice $U = X + WS$ made in the p.d. case. Here $m \times t$ matrix $W$ is the inflation factor. Note that this choice of $U$ is still a generalization of Costa's (original) choice.

Using the capacity formula of [7], we obtain the achievable rate as given by

$$R = \mathbb{E}_{\hat{H}}\left(\mathbb{E}_{H|\hat{H}}\log\{|H(TT^* + \Sigma_S)H^* + \Sigma_Z|\} - \quad (1)\right.$$

$$\left.\min_W \mathbb{E}_{H|\hat{H}}\log\begin{vmatrix} I_m + W\Sigma_S W^* & (T^* + W\Sigma_S)H^* \\ H(T + \Sigma_S W^*) & H(TT^* + \Sigma_S)H^* + \Sigma_Z \end{vmatrix}\right).$$

Under this choice of $U$, the perfect-CSIT optimal inflation factor is given by $W_{opt} = T^*H^*(HTT^*H^* + \Sigma_Z)^{-1}H$.

We define the no-interference upper-bound $C$ as the rate achievable over the FDPC in absence of interference, i.e., when $Q = 0$. Thus if the covariance matrix of $\Sigma_X$ is used over the FDPC, then $C = \mathbb{E}_H \log \frac{|\Sigma_Z + H\Sigma_X H^*|}{|\Sigma_Z|}$.

## III. Determination of Inflation Factor

The minimization in equation (1) is the problem of determination of inflation factor. As noted in [2], it is a non-convex optimization problem and it seems intractable to find an analytical closed-form solution for the optimal inflation factor. We generalize below our previously-developed algorithms to apply to the problem at hand.

*Algorithm 1:* In this algorithm[1], we minimize the objective function stepwise, i.e., at each step, we minimize over one row of $W$ while treating all other rows as constants. Let us first consider the minimization over $W_{(1)}$. Similar to the minimization problem considered in [2], in this problem also, only the first row and the first column of the block-partitioned matrix in (1), which we call $M$, depend on $W_{(1)}$. Therefore, we repartition $M$ as $\begin{bmatrix} a & B^* \\ B & D \end{bmatrix}$, where $a = M_{(11)}$, $B^* = [W_{(1)}\Sigma_S(W^*)^{(\bar{1})} \ (T^*_{(1)} + W_{(1)}\Sigma_S)H^*]$, and $D$ is $(m+r-1) \times (m+r-1)$ matrix remained after excluding the first row and the first column from $M$. Now $|M| = |D|(a - B^*D^{-1}B)$. Clearly, the problem reduces to the minimization of $E_H \log(a - B^*D^{-1}B)$ over $W_{(1)}$. Towards this end, observe that the expression $(a - B^*D^{-1}B)$ is a quadratic in $W_{(1)}$. We now upper-bound the objective function through Jensen's Inequality by moving the expectation inside the logarithm, then the technique of 'completing the square' can be used to minimize over $W_{(1)}$.

If $m = 1$, the optimal $W$ is given by equation (2) at the bottom of the next page, otherwise proceed as follows. Partition $D^{-1}$ as $\begin{bmatrix} F & G \\ J & K \end{bmatrix}$, where $F$ is $(m-1) \times (m-1)$ square matrix, $K$ is $r \times r$ matrix, and $G^* = J$. Then we can obtain $W_{(1)}$ as given by equation (3)[2]. The required expectations can be evaluated numerically.

To minimize over the $k^{th}$ row of $W$, a similar idea can be used. Thus, an iterative algorithm can now be set up,

as done in [2], in which one iteration consists of step-wise minimizations over all the rows of $W$ and these iterations are repeated until a good choice is found. Importantly, this algorithm converges [2].

*Algorithm 2:* In this algorithm, we find a stationary point of the objective function, i.e., solve for Karush-Kuhn-Tucker conditions of the optimization. Thus we need to solve an equation $\frac{d}{dW}\mathbb{E}_{H|\hat{H}}\log|M| = 0$. It can be proved, as done in [2], that the above equation reduces to $\mathbb{E}_{H|\hat{H}}(A_1 W + A_2^* H)\Sigma_S = 0$, where $[A_1 \ A_2^*]^* = M^{-1}[I_m \ 0]^*$. Without loss of generality (i.e., even if $|\Sigma_S| = 0$), we can consider a solution of the form $W = -(\mathbb{E}_{H|\hat{H}}A_1)^{-1}\mathbb{E}_{H|\hat{H}}(A_2^*H) = g(W)$ for the above equation[3]. Using this solution, an iterative numerical algorithm can now be set up in which at the $n^{th}$ iteration, we set $W^{(n)} = g(W^{(n-1)})$.

## IV. High and Low SNR Analysis

### A. High SNR Scaling Factor

The scaling factor of $\min(t,r)$ is achievable over the no-CSIT FDPC, irrespective of $\Sigma_S$, when $\Sigma_X$ is p.d. and it is equal to that of the no-interference upper-bound [2]. In the p.s.d. case, the high-SNR scaling factor may not necessarily be equal to that of the no-interference upper-bound and it also depends on $\Sigma_S$. Interestingly, such characteristics are never observed in the p.d. case.

*Theorem 1:* Over the no (and hence, partial) CSIT FDPC with p.s.d. input covariance matrix, the largest DPC-achievable[4] scaling factor is given by $\min\{r, \mathrm{rank}(\Sigma_X + \Sigma_S)\} - \min\{r, \mathrm{rank}(\Sigma_X + \Sigma_S) - m\}$, provided the ratio $\frac{Q}{P}$ is held constant as $P \to \infty$, and the fading process $H$ is such that for any p.s.d. $\Sigma$, $\mathrm{rank}(H\Sigma H^*) = \min(r, \mathrm{rank}(\Sigma))$ with probability 1.

*Proof:* Let us reformulate the achievable-rate expression as follows: define $T$ via $\Sigma_X = PTT^*$ so that $X' \sim \mathcal{CN}(0, PI_m)$. Thus, we obtain $R$ as given by equation (4) at the bottom of the next page where $\Sigma_S = P\Sigma'_S$. In order to maximize the achievable scaling factor, we need to choose $W$ such that 1) $\log|P(I_m + W\Sigma'_S W^*)|$ does not achieve the high-SNR scaling factor larger than $m$, and 2) the term $\mathbb{E}_H \log|\Sigma_Z + PHAH^*|$, where $A = TT^* + \Sigma'_S - (T + \Sigma'_S W^*)(I_m + W\Sigma'_S W^*)^{-1}(T^* + W\Sigma'_S)$, scales minimally in the high-SNR regime.

Let us analyze the second condition. It can be proved that $A$ is p.s.d. for any choice of $W$. This and the assumption made about $H$ imply that the term $\mathbb{E}_H \log|\Sigma_Z + PHAH^*|$ would scale at high-SNR as $\min(r, \mathrm{rank}(A))\log\mathrm{SNR}$. Therefore we need to minimize $\mathrm{rank}(A)$ over $W$. To this end, we prove that $\mathrm{rank}(A) \geq \mathrm{rank}(TT^* + \Sigma'_S) - m$. This lower-bound on $\mathrm{rank}(A)$ is achievable with the choice $W = T^+$. Also, this choice of $W$ satisfies the first condition. ∎

---

[1] Notations used in this algorithm: For any matrix $A$, $A_{(1)}$ denotes the first row of it; $A^{(1)}$ its first column while $A_{(11)}$ is its $(1,1)$ element; $A_{(\bar{1})}$ and $A^{(\bar{1})}$ denote entire matrix $A$ except for its first row and except for its first column, respectively.

[2] Certain matrices inverted in equations (2) and (3) are singular if and only if $\Sigma_S$ is p.s.d. This can be fixed by writing $\Sigma_S = T_2 T_2^*$, similar to what is done when $\Sigma_X$ is p.s.d.

[3] Matrix $A_1$ is invertible with probability 1.

[4] The high-SNR scaling factor derived in Theorem 1 is optimal under the choice of DPC-based transmission strategy, but need not be the best achievable under all transmission strategies. This is because Costa's choice of auxiliary RV, which is considered here, is not known to be optimal under imperfect CSIT.

The assumption regarding the fading process $H$ made above warrants a discussion. This assumption is fairly general and many fading distributions of practical interest would fall in this category. For example, if the probability that null($H$) contains certain fixed subspace (of the t-dimensional space) is zero then such a fading distribution would fall in the category, and the commonly studied Rayleigh fading satisfies this condition. If this assumption is not made, the scaling factor would turn out to be dependent on the fading distribution.

Let us now discuss the cases when the scaling factor of the no-CSIT FDPC and that of the no-interference upper-bound are equal. When $t \leq r$, two are equal (to $m$), irrespective of the choices of $\Sigma_X$ and $\Sigma_S$. However, unlike the case of p.d. covariance matrix [2], now, under p.s.d. input covariance, the difference between the no-interference upper-bound and the achievable rate may not necessarily go to zero in the limit of high SNR. Now consider $t > r$. If $t > r \geq \text{rank}(\Sigma_X + \Sigma_S)$, two scaling factors are equal. On the other hand, when $t \geq \text{rank}(\Sigma_X + \Sigma_S) > r$, two are equal only if $\text{rank}(\Sigma_X + \Sigma_S) = m$.

It is worth mentioning here that whenever $r \geq m = \text{rank}(\Sigma_X + \Sigma_S)$, it can be shown that the difference between the no-interference upper-bound and the achievable rate goes to zero in the limit as $P \to \infty$. This follows from Theorem 2 of [2].

Note that Proposition 3 in [2] about the DPC-achievable scaling factor of the no-CSIT Gaussian MIMO BC relies on the use of p.d. input covariance matrix for the last user. If a p.s.d. covariance matrix is used instead, we may get a scaling factor lesser than $\min(t,r)$.

### B. Low SNR Analysis

*Theorem 2:* With the choice of $W = 0$, the ratio of the rate achievable over the no-CSIT FDPC to the no-interference upper-bound goes to 1 in the limit of low SNR if the ratio $\frac{Q}{P}$ is kept constant.

*Proof:* With $W = 0$, we have
$R = \mathbb{E}_H \log |\Sigma_Z + H(\Sigma_X + \Sigma_S)H^*| - \mathbb{E}_H \log |\Sigma_Z + H\Sigma_S H^*|$;
and the no-interference upper-bound as

$C = \mathbb{E}_H \log |\Sigma_Z + H\Sigma_X H^*| - \log |\Sigma_Z|$.

Taylor-series expansions of $R$ and $C$ about $P = 0$ would be of the form $(R,C) = \alpha_{(R,C)} P + \beta_{(R,C)} P^2 + o(P^2)$, where $\alpha_{(R,C)} = \frac{d(R,C)}{dP}\big|_{P=0}$, and $\beta_{(R,C)} = \frac{d^2(R,C)}{dP^2}\big|_{P=0}$. Hence, if $\alpha_R = \alpha_C$, we must have $\lim_{P \to 0} \frac{R}{C} = 1$.

Let us first derive $\alpha_R$. Let $\Sigma_X = P\Sigma'_X$, $\Sigma_S = P\Sigma'_S$, $[\Sigma_Z]_{ij} = z_{ij}$, $[H\Sigma'_X H^*]_{ij} = a_{ij}$, and $[H\Sigma'_S H^*]_{ij} = b_{ij}$. Then for $M = \Sigma_Z + PH(\Sigma'_X + \Sigma'_S)H^*$, we have

$$\frac{d}{dP}\log|M| = \frac{1}{|M|}\sum_{i=1}^r |M^{(i)}|, \text{ where}$$
$$[M^{(i)}]_{jk} = z_{jk} + P(a_{jk} + b_{jk}) \cdots \text{ if } j \neq i, \text{ and}$$
$$= (a_{jk} + b_{jk}) \cdots \text{ if } j = i.$$

Here, $M^{(i)}$ is nothing but the matrix $M$ with $i^{\text{th}}$ row replaced by its derivative with respect to $P$. This step can be verified from the so-called 'ugly' formula of determinants.

Letting $N = \Sigma_Z + PH\Sigma'_S H^*$, we get $\alpha_R =$

$$\mathbb{E}_H \frac{1}{|\Sigma_Z|}\sum_{i=1}^r \left\{\left(|M^{(i)}| - |N^{(i)}|\right)\Big|_{P=0}\right\} = \mathbb{E}_H \frac{1}{|\Sigma_Z|}\sum_{i=1}^r |Q^{(i)}|,$$
where $[Q^{(i)}]_{jk} = z_{jk} \cdots \text{ if } i \neq j, \text{ and}$
$$= a_{jk} \cdots \text{ if } i = j.$$

This follows because at $P = 0$, all rows of $M^{(i)}$ and those of $N^{(i)}$ are equal except the $i^{\text{th}}$ row. The result then follows by noting that $a_{jk} = [H\Sigma'_X H^*]_{jk}$. ∎

Thus, in the limit of low SNR, it is optimal, in a 'ratio' sense, to treat the interference as noise. Note that this result holds irrespective of whether p.d. or p.s.d. covariance matrix is used.

### C. High-SNR Analysis: p.d. covariance matrix

We include in this subsection one important result regarding the p.d. case. It is proved in [2] that the choice $W = I_t$ is optimal in the high-SNR regime if $t \leq r$. We now prove a similar result for the remaining case of $t > r$.

$$W = T^*\mathbb{E}\{H^*[H(TT^* + \Sigma_S)H^* + \Sigma_Z]^{-1}H\}\Sigma_S(\Sigma_S - \Sigma_S\mathbb{E}\{H^*(H(TT^* + \Sigma_S)H^* + \Sigma_Z)^{-1}H\}\Sigma_S)^{-1}, \quad (2)$$

$$W_{(1)} = \left((I_m)_{(1)}T^*\mathbb{E}(H^*J)W_{(\bar{1})}\Sigma_S + (I_m)_{(1)}T^*\mathbb{E}(H^*KH)\Sigma_S\right) \times \left(\Sigma_S - \Sigma_S(W_{(\bar{1})})^*\mathbb{E}(F)W_{(\bar{1})}\Sigma_S\right.$$
$$\left. -\Sigma_S\mathbb{E}(H^*J)W_{(\bar{1})}\Sigma_S - \Sigma_S(W_{(\bar{1})})^*\mathbb{E}(GH)\Sigma_S - \Sigma_S\mathbb{E}(H^*KH)\Sigma_S\right)^{-1}, \quad \cdots \text{ where } \mathbb{E}(\cdot) \equiv \mathbb{E}_{H|\hat{H}}(\cdot). \quad (3)$$

$$R = \mathbb{E}_H \log \frac{|PI_m||\Sigma_Z + PH(TT^* + \Sigma'_S)H^*|}{|P(I_m + W\Sigma'_S W^*)||\Sigma_Z + PH\{TT^* + \Sigma'_S - (T + \Sigma'_S W^*)(I_m + W\Sigma'_S W^*)^{-1}(T^* + W\Sigma'_S)\}H^*|}. \quad (4)$$

$$R = \mathbb{E}_H \log \frac{|\Sigma_X||\Sigma_Z + H(\Sigma_X + \Sigma_S)H^*|}{|\Sigma_X + W\Sigma_S W^*||\Sigma_Z + PH\{\Sigma'_X + \Sigma'_S - (\Sigma'_X + \Sigma'_S W^*)(\Sigma'_X + W\Sigma'_S W^*)^{-1}(\Sigma'_X + W\Sigma'_S)\}H^*|}. \quad (5)$$

4*Theorem 3:* For the no-CSIT FDPC with $t > r$ and p.d. input covariance matrix, the inflation factor optimal[5] in the high-SNR regime is given by $W = I_t + \triangle W$, where $\triangle W$ is any matrix such that $\triangle W \Sigma_S = 0$, provided the ratio $\frac{Q}{P}$ is kept constant, and $H$ is such that for any p.s.d. $\Sigma$, $\text{rank}(H\Sigma H^*) = \min(r, \text{rank}(\Sigma))$ with probability $> 0$.

*Proof:* The achievable-rate expression from [2] can be rewritten to get equation (5) at the bottom of the previous page. Let $A(W) = \Sigma'_X + \Sigma'_S - (\Sigma'_X + \Sigma'_S W^*)(\Sigma'_X + W\Sigma'_S W^*)^{-1}(\Sigma'_X + W\Sigma'_S)$. The idea is to characterize the set of all choices of $W$ that achieve the optimal scaling. It turns out that the choices of W described in the statement of the theorem are the only choices that achieve the optimal scaling, and hence follows the theorem.

---
[5] Since we considering the optimality of $W$, let $W_P$ be describe the inflation factor used at power P. We assume in this theorem that $\lim_{P \to \infty} W_P$ exists.

Now, $R$ scales optimally if and only if (iff) $W$ is such that 1) the factor $\log |\Sigma_X + W\Sigma_S W^*|$ does not achieve the high-SNR scaling factor larger than $t$, and 2) the term $\mathbb{E}_H \log |\Sigma_Z + PHA(W)H^*|$ does not scale in the high-SNR regime. To show the dependence of $A$ on the transmitted power $P$, let us write $A(W)$ as $A(W_P)$. Note that $A(W_P)$ is p.s.d. for any choice of $W_P$.

Let us analyze the second condition. The assumption made regarding $H$ implies that the second condition holds iff $\lim_{P \to \infty} \text{tr}\{A(W_P)\} = 0$. Since the matrix inverse is a continuous function (note $(\Sigma'_X + W_P \Sigma'_S W_P^*)$ is always p.d.) and we have assumed that the limit of $W_P$ exists, the above implies the existence of $W = W_\infty = \lim_{P \to \infty} W_P$ such that $\text{tr}\{A(W)\} = \text{tr}\{A(W_\infty)\} = 0$. Since $A(W)$ is p.s.d. for any $W$, this implies $A(W) = 0$. Therefore, let us characterize the set of all $W$'s for which $A(W)$ can be zero.

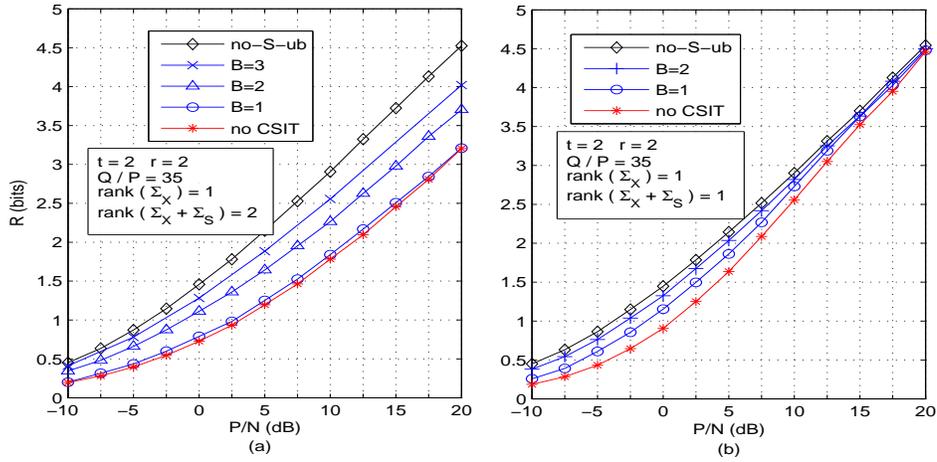

Fig. 1
ACHIEVABLE RATE VS. SNR FOR $2 \times 2$ SYSTEM. (A) $\text{rank}(\Sigma_X + \Sigma_S) > \text{rank}(\Sigma_X)$. (B) $\text{rank}(\Sigma_X + \Sigma_S) = \text{rank}(\Sigma_X)$.

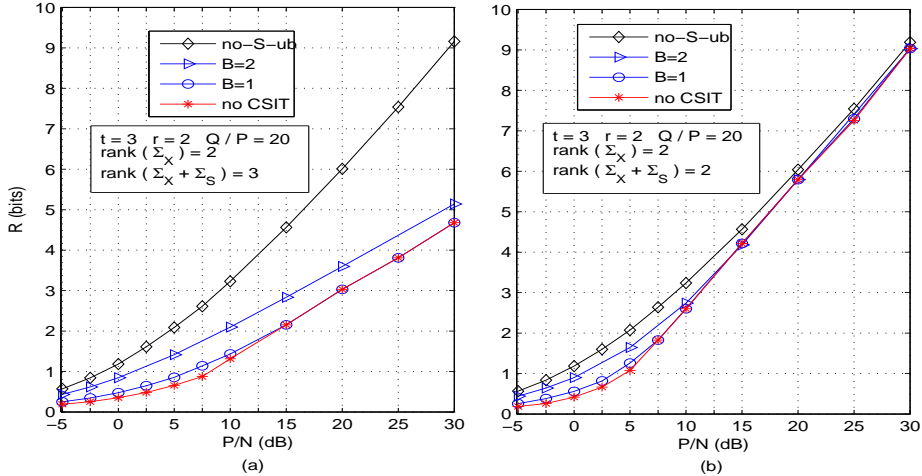

Fig. 2
ACHIEVABLE RATE VS. SNR FOR $3 \times 2$ SYSTEM. (A) $\text{rank}(\Sigma_X + \Sigma_S) > \text{rank}(\Sigma_X)$. (B) $\text{rank}(\Sigma_X + \Sigma_S) = \text{rank}(\Sigma_X)$.

At optimal $W$, for which $A(W) = 0$, the matrix $(\Sigma'_X + W\Sigma'_S)$ must be invertible since matrices $(\Sigma'_X + \Sigma'_S)$ and $(\Sigma'_X + W\Sigma'_S W^*)$ are p.d. Therefore at optimal $W$, $\Sigma'_X + W\Sigma'_S W^* - (\Sigma'_X + W\Sigma'_S)(\Sigma'_X + \Sigma'_S)^{-1}(\Sigma'_X + \Sigma'_S W^*) = 0$. The left hand side of this equation is a quadratic in $W$. Hence, it can be written as $(Wa - b)(Wa - b)^* + c$ for some appropriately chosen $a$, $b$, and $c$. Equating terms, we get $aa^* = \Sigma_S - \Sigma_S(\Sigma_X + \Sigma_S)^{-1}\Sigma_S$ and $ab^* = \Sigma_S(\Sigma_X + \Sigma_S)^{-1}\Sigma_X$. Note that $aa^* = ab^*$ and hence without loss of generality, we can set $a = b$. This implies $c = 0$.

Case I: $\Sigma_S$ is p.d. $\Leftrightarrow aa^*$ is p.d. Then there exists a unique solution $W = I_t$.

Case II: $\Sigma_S$ is p.s.d. $\Leftrightarrow aa^*$ is p.s.d. In this case, $W = I_t$ is one of the possibly many solutions. Let the optimal solution be $W = I_t + \triangle W$. Then, we must have $(\triangle W)aa^*(\triangle W)^* = 0$. But $\text{rank}(aa^*) = \text{rank}(\Sigma_S)$, and hence $\text{null}(aa^*) = \text{null}(\Sigma_S)$. This implies $\triangle W \Sigma_S = 0$. ∎

Since $W$ always gets multiplied by $\Sigma_S$ in the rate expression, we can set $\triangle W = 0$, without loss of generality.

## V. Numerical Results

In Figs. 1 to 3, we consider real-valued FDPCs. Here, the elements of $H$ are i.i.d. $\sim \mathcal{N}(0,1)$. We quantize each element of $H$ separately using a simple 'equally spaced level' quantizer (quantization bins are of equal length except for the first and the last bins which extend to infinity) and the spacing is determined using data from [8]. $B$ denotes number of feedback bits per element of $H$. 'no-S-ub' denotes the no-interference upper-bound. Also in these results, we use Algorithm 1. Figs. 1 and 2 have two parts. The only difference between the FDPCs of part (a) and (b) is that the choice of $\Sigma_S$ is varied [6].

In Figs. 1 and 2, we consider $2 \times 2$ and $3 \times 2$ FDPCs, respectively. In both figures, we see a significant increase in the achievable rate with increasing feedback. Observe that for the $2 \times 2$ FDPC, the high-SNR scaling factor of $R$ and that of the 'no-S-ub' are equal whereas for the $3 \times 2$ FDPC of Fig. 2, the scaling factor of $R$ is dependent on the choice of $\Sigma_S$. Interestingly, in Figs. 1(b) and 2(b), the difference between the no-interference upper-bound and the achievable rate can be seen to tend to zero with increasing SNR. The high-SNR behavior of the achievable rate observed in these figures is in accordance with the analytical results proved earlier.

In Fig. 3, we consider the low-SNR behavior. At sufficiently low SNR, the choice $W = 0$ can be said to be near optimal. This fact is formulated mathematically and proved in Theorem 2.

In Fig. 4, we present comparison of two algorithms. We take complex-valued systems with $H$ Rayleigh faded and also consider the case of no CSIT. It is observed rather surprisingly that these algorithms typically yield $W$'s that achieve almost equal rate and therefore, we present here some cases of difference. For FDPC 1, $m = 1$ and Algorithm 1 uses the closed-form solution obtained in equation (2). Curves for FDPC 1 suggest that Algorithm 2 yields a solution that is slightly better than this closed-form solution of (2). The main advantage of Algorithm 1 over the second is that it is faster in terms of number of iterations required.

[6] All parameters necessary to generate the plots enclosed in this paper can be obtained by e-mailing the first author.

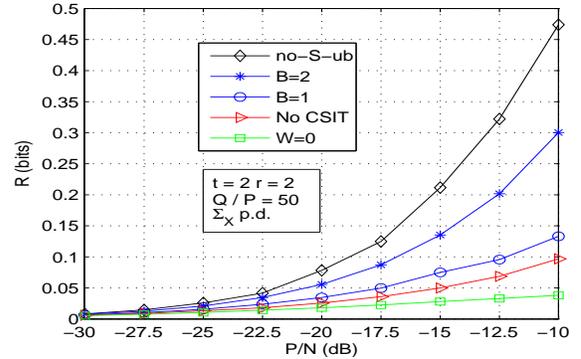

Fig. 3
Low-SNR Analysis: Achievable Rates vs. SNR.

## VI. Joint Optimization over the Input Covariance Matrix and the Inflation Factor

We solve this problem of joint optimization under an additional constraint on $\text{rank}(\Sigma_X)$.

Since using Algorithms 1 and 2, we can determine $W$ for a given choice of $\Sigma_X$, let us first consider optimizing over $\Sigma_X$ for a given $W$. Let the rank of $\Sigma_X$ be constrained by $m$. Under this condition, we write $\Sigma_X = TT^*$ for some $t \times m$ matrix $T$, and therefore, $\text{rank}(\Sigma_X) \leq m$. This also ensures that $\Sigma_X$ is p.s.d. Thus we have the following optimization problem: $\max_T R$, subject to the power constraint $\text{tr}(TT^*) \leq P$.

To derive the necessary conditions for the optimization, we form the lagrangian $J$ and set $\frac{\partial J}{\partial T} = 0$. We have

$$J = \mathbb{E}_H \log \frac{|\Sigma_Z + H(TT^* + \Sigma_S)H^*|}{\begin{vmatrix} I_m + W\Sigma_S W^* & (T^* + W\Sigma_S)H^* \\ H(T + \Sigma_S W^*) & \Sigma_Z + H(TT^* + \Sigma_S)H^* \end{vmatrix}} - \lambda \text{tr}(TT^*),$$

where $\lambda$ is the lagrange multiplier.

As mentioned before, the problem of covariance optimization for the MIMO channel has been studied in [4], [5], [6]. In [4], a suboptimal solution is provided; whereas in [5] and [6], the authors solve for the Karush-Kuhn-Tucker conditions, and to this end, differentiate the lagrangian with respect to each element of $T$ separately. This technique would be very cumbersome for the problem at hand. The key here is to differentiate the lagrangian with respect to matrix $T$ directly using the techniques of matrix differentiation. Due to lack of space, we omit the details of finding derivatives and directly state below the necessary conditions for the above optimization:

$$\lambda T = \mathbb{E}_H \left\{ H^* N_r^{-1} HT + [0 \ H^*] D_r^{-1} \begin{bmatrix} I_m \\ HT \end{bmatrix} \right\} = g(T, W).$$

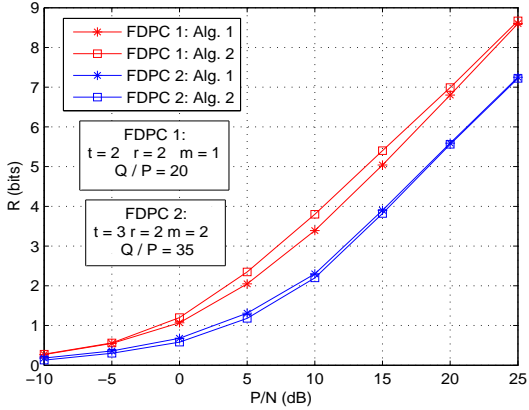

Fig. 4
COMPARISON OF ALGORITHMS: ACHIEVABLE RATES VS. SNR.

This fixed-point equation allows us to set up an iterative algorithm for the joint optimization.
1. Start with some initial choice for $T$, say $T^{(0)}$.
2. At the $n^{th}$ iteration:
   - For the choice $T^{(n-1)}$ of $T$, find $W$.
   - For $W$ obtained above, find optimal $T$. Towards this end, we use the above equation, and set $T^{(n)} = \frac{1}{\lambda} g(T^{(n-1)}, W)$, where the required expectations are computed numerically and the lagrange multiplier $\lambda$ is evaluated so as to meet the power constraint.
3. Repeat the above step until the increase in the achievable rate is negligible.

<u>*Numerical Results:*</u> In Fig. 5, $FDPC$ is the rate achievable over the FDPC with the jointly-optimized input covariance and the inflation factor and similarly for curves labeled $MIMO$. $FDPC_{lb}$ is the rate achievable using the scaled-version of identity as the input covariance matrix and the optimal inflation factor (determined using Algorithm 1). If $x$ is the ratio of the largest to the smallest eigen-value of the input covariance matrix (obtained using the algorithm of joint optimization), then in Fig. 5(b), 'Eig. Ratio' denotes $\frac{1}{3}\log_{10}(x)$.

In Fig. 5(a), we consider correlated Rayleigh fading with separable correlations, whereas in Fig. 5(b), we take elements of $H$ to be i.i.d. $\sim \mathrm{Unif}[0,1] + j\mathrm{Unif}[0,1]$ where $\mathrm{Unif}[0,1]$ denotes uniform distribution over $[0,1]$. In these figures, the improvement due to spatial water-filling is evident. In Fig. 5(c), we consider a rank-constrained optimization and $H$ is Rayleigh faded. One can see the sufficiency of reduced-rank signalling at low SNR (rank 1 at $-10$ dB, 2 at $-5$ dB) while the full-rank signalling performs significantly better (than the reduced-rank signalling) at high SNR of 30 dB.

## VII. CONCLUSION

The determination of the inflation factor under partial CSIT was the major obstacle in the problem of applying DPC to the partial-CSIT Gaussian MIMO BC. The results of this and our earlier paper [2] make significant progress to address this problem. At this time, the problem of jointly optimizing the sum-rate of the partial-CSIT Gaussian MIMO BC over the covariance matrices of all the users (and the inflation factors) seems to be the next logical step in developing an algorithmic solution for computing good lower bounds on the achievable sum-rate of the Gaussian MIMO BC.

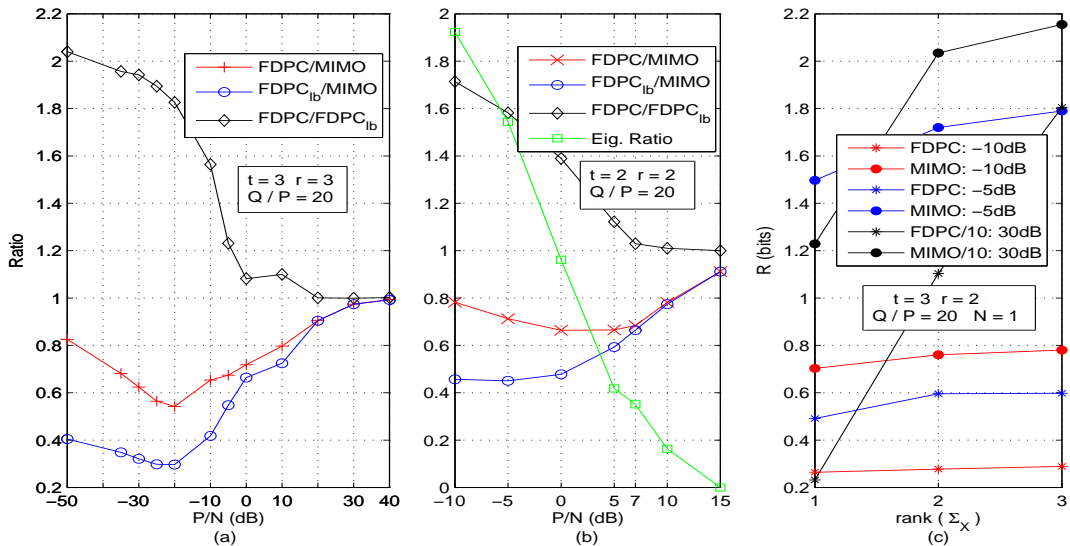

Fig. 5
COVARIANCE OPTIMIZATION. (A) RATIO OF VARIOUS RATE PAIRS FOR $3 \times 3$ SYSTEM. (B) RATIOS FOR $2 \times 2$ SYSTEM. (C) RANK CONSTRAINED OPTIMIZATION FOR $3 \times 2$ SYSTEM.